\documentclass[fleqn,10pt]{wlscirep}
\usepackage[utf8]{inputenc}
\usepackage[T1]{fontenc}
\usepackage{braket}
\usepackage{multirow}
\definecolor{lightblue}{rgb}{0.9,0.9,1}
\definecolor{lightred}{rgb}{1,0.9,0.9}
\definecolor{lightgreen}{rgb}{0.9,1,0.9}
\usepackage{tabularx}
\usepackage{footnote}
\usepackage{tablefootnote}
\usepackage{float}
\usepackage{graphicx}
\usepackage{amsmath}
\usepackage{graphicx,wrapfig,lipsum}
\usepackage{xcolor}
\usepackage{colortbl}
\makeatletter
\newcommand*\bigcdot{\mathpalette\bigcdot@{.5}}
\newcommand*\bigcdot@[2]{\mathbin{\vcenter{\hbox{\scalebox{#2}{$\m@th#1\bullet$}}}}}
\makeatother
\definecolor{lightblue}{rgb}{0.88,0.96,1}
\definecolor{lightgray}{rgb}{0.9,0.9,0.9}
\usepackage{cite}
\usepackage{multirow}
\usepackage{hyperref}
\usepackage{amsfonts}
\usepackage{bbold}
\usepackage{subcaption}
\title{Indispensability of orbital angular momentum states in secure quantum communication tasks}
\author[1*]{Rajni Bala}
\author[1]{Sooryansh Asthana}
\affil[1]{Department of Physics, Indian Institute of Technology Delhi, New Delhi-110016, India}

\affil[*]{Rajni.Bala@physics.iitd.ac.in}

\begin{abstract}
 Quantum key distribution protocols have been designed for layered networks employing multidimensional entangled and separable orbital angular momentum states  \cite{pivoluska2018layered, bala2023quantum}. This paper seeks an answer to the overarching question-- in the context of secure quantum communication tasks, do orbital angular momentum states act merely as an alternative or do they act as an indispensable resource? We start by showing that the task of quantum key distribution in layered networks can also be accomplished with several copies of lower-dimensional states such as polarization qubits. For this reason, orbital angular momentum states do not offer any intrinsic advantage in layered quantum key distribution. The potential of OAM states unveils itself in the enhancement of key generation rates and integrated quantum communication tasks, which we present in this paper. These tasks can be implemented exclusively with high-dimensional OAM entangled states.  In fact, we show that the employment of orbital angular momentum states eliminates the need for entangled state measurements, whose implementation is resource-intensive.  We believe that this study opens up a possibility for designing several quantum information processing tasks in which multidimensional OAM states act as an indispensable resource.
\end{abstract}
\begin{document}

\flushbottom
\maketitle

\thispagestyle{empty}

\section*{Introduction}
\label{intro}
Orbital angular momentum (OAM) states of light, characterized by their helical phase fronts \cite{allen1992orbital, mair2001entanglement}, have emerged as promising candidates for high-dimensional photonic qudits since they theoretically belong to an infinite-dimensional Hilbert space. In fact, there is a significant advancement in the generation and measurement of single photon and entangled OAM states \cite{erhard2018twisted,willner2021orbital,d2013test}. This has led to proposals for a number of secure communication protocols with OAM states,  including layered quantum key distribution protocols and hierarchical secret sharing \cite{mirhosseini2015high, mafu2013higher,  pivoluska2018layered, bala2023quantum, sit2017high,hillery1999quantum,qin2020hierarchical,sekga2023measurement}. Recall that a layered network consists of a number of participants divided into different overlapping subsets, termed as {\it layers}.
Though the employment of multi-dimensional OAM states offers the advantage of realizing these tasks simultaneously in all the layers, these states are merely an alternative for accomplishing these tasks. This is because multidimensional entangled states fall into two categories-- reducible entangled states and irreducible entangled states \cite{kraft2018characterizing,cong2017witnessing}. Reducible entangled states are states that can be written as a tensor product of multiqubit states. As an example, consider the state employed in \cite{pivoluska2018layered} as a resource state for layered quantum key distribution, 
\begin{align}
    |\Psi\rangle\equiv\frac{1}{2}\Big\{\Big(|00\rangle+|22\rangle\Big)|0\rangle+\Big(|11\rangle+|33\rangle\Big)|1\rangle\Big\}_{AB_1B_2}.
\end{align}
Under the decimal-to-binary mapping ($|0\rangle\equiv |00\rangle, |1\rangle\equiv |01\rangle, |2\rangle\equiv |10\rangle, |3\rangle\equiv |11\rangle$), the state $|\Psi\rangle$ can be reexpressed in the form,
\begin{align}
|\Psi\rangle\equiv \frac{1}{{2}}\Big(|00\rangle+|11\rangle\Big)_{AB_1}\Big(|000\rangle+|111\rangle\Big)_{AB_1B_2}.
\end{align}
In contrast, irreducible entangled states can not be written as a tensor product of lower-dimensional entangled states, e.g., 
\begin{align}
|\Phi\rangle\equiv \frac{1}{2}\Big\{\Big(|00\rangle+|22\rangle\Big)|0\rangle+\Big(|11\rangle-|33\rangle\Big)|1\rangle\Big\}_{AB_1B_2}.
\end{align}

Since reducible entangled states are employed as resources for quantum key distribution in layered networks \cite{pivoluska2018layered}, the same task can also be realized by sequentially distributing multiple copies of polarization qubits. To demonstrate this, in this paper, we first show how the task of simultaneous key distribution in a layered network can be implemented with both multi-dimensional OAM states and sequential sharing of  multi-copies of polarized qubits. We also present the similarities and contrasts between the two protocols. Notably, if we want to enhance the key generation rate in a given layer, OAM states become an indispensable choice.

This motivates us to ask the question, which is the main thrust of this paper: {\it do there exist secure quantum communication tasks in which orbital angular momentum states offer an intrinsic advantage?} To answer this, we move on to identify a class of tasks that would intrinsically require OAM states for their implementations. These tasks employ those high-dimensional entangled states that cannot be simulated by employing sequential multicopies of lower-dimensional systems, i.e., these tasks employ irreducible entangled states \cite{kraft2018characterizing,guo2020measurement,sun2020experimental}. In this manner, these states bring out the efficacy of multidimensional entangled OAM states as compared to multiqubit polarized states. In the first task, we show how such states allow one to have a controlled distribution of a key as well as a secret depending solely on the will of the controller. In the second task, we show how the employment of irreducible multidimensional entangled OAM states allows for the basis-dependent implementation of various tasks. This, in turn, utilizes the full potential of resource states, thus minimizing the data that is otherwise discarded. These tasks, of course, due to the employment of irreducible multidimensional entangled OAM states,  cannot be simulated with sequential multi copies of polarized entangled qubits, thereby showing the intrinsic advantage offered by OAM states of light in secure communication.  To the best of our knowledge, these protocols are the first applications of irreducible entangled OAM states.

The plan of the paper is as follows: in section (``\hyperlink{Quantum key distribution in layered networks: Existing protocols}{Quantum key distribution in layered networks: Implementable with both-- multidimensional  OAM states and multiple copies of polarisation entangled states}''), we briefly recapitulate four different protocols for distributing secure keys in a network of three participants distributed in two layers.  Section (``\hyperlink{Tasks implementable with irreducible multidimensional entangled OAM states only}{Tasks implementable with irreducible multidimensional entangled OAM states only}'') is central to the paper, in which we present two secure quantum communication tasks-- (i) \hyperlink{controlled-secret sharing and key distribution}{controlled-secret sharing and key distribution}, (ii) \hyperlink{Basis-dependent 
 simultaneous secret sharing and key distribution}{Basis-dependent simultaneous secret sharing and key distribution}-- that can be  implemented with multidimensional entangled OAM states only. Section (``\hyperlink{Conclusions}{Conclusions}'') concludes the paper.

\section*{Layered quantum key distribution: Implementable with both-- multidimensional OAM states and multiple copies of polarization-entangled states}
\label{formalism}
\hypertarget{Quantum key distribution in layered networks: Existing protocols}{}

In this section, we present illustrative protocols for layered quantum key distribution by employing both--reducible multidimensional OAM states and entangled multiqubit polarization states in tables (\ref{OAM}) and (\ref{qubit}) respectively.  In layered networks, participants are divided into specific subsets, termed as {\it layers}.  Three protocols have been proposed for QKD in layered networks, by employing (i) entangled multidimensional states \cite{pivoluska2018layered}, (ii) separable multidimensional states \cite{bala2023quantum}, (iii) entangled multiqubit states \cite{pivoluska2018layered}.  
\begin{table}[h!]
\centering
\begin{center}
\resizebox{!}{6cm}{
\begin{tabular}{|c|c|c|}
\hline

\multirow{2}{*}{\textbf{Task}} &\multicolumn{2}{c|}{\multirow{2}{*}{\textbf{Key distribution in layered networks}}}\\
&\multicolumn{2}{c|}{}\\
\hline

\multirow{4}{*}{\textbf{Features $\downarrow$}} &&\\
&\textbf{Protocol I} & \textbf{Protocol II}\\
& \multirow{1}{*}{\textbf{(with multidimensional entangled OAM states)}} & \multirow{1}{*}{\textbf{(with multidimensional separable OAM states)}}\\
&&\\
\hline

\multirow{3}{*}{\textbf{Resource states}} & \multirow{3}{*}{$\ket{\Psi}= \frac{1}{2}\Big\{\Big(\ket{00}+\ket{22}\Big)\ket{0}+\Big(\ket{11}+\ket{33}\Big)\ket{1}\Big\} $} & \multirow{2}{*}{$S_1  : \{|00\rangle,|11\rangle,|20\rangle,|31\rangle\}$}\\
&&\\
&&$S_2  : \{|0' + \rangle,|1' - \rangle,|2' + \rangle,|3' - \rangle\}$\\
\hline
\multirow{4}{3cm}{\textbf{Initialisation by Alice}} &\multirow{4}{6cm}{Keeps the first subsystem with herself. Sends the second and third subsystems to Bob\(_1\) and Bob\(_2\) respectively.} & \multirow{4}{6cm}{ Randomly prepares a state from \(S_1\) or \(S_2\) and sends the first and the second subsystems to Bob\(_1\) and Bob\(_2\) respectively.}\\
&&\\
&&\\
&&\\
\hline

\multirow{4}{3cm}{\textbf{Operations}} &\multirow{3}{6cm}{Measurement on the received systems either in the computational (OAM) or Fourier (angular) basis.} & \multirow{3}{6cm}{Measurement on the received systems either in the computational (OAM) or Fourier basis (angular).}\\
&&\\
&&\\
&repeated for a large number of rounds& repeated for a large number of rounds\\
\hline

\multirow{4}{3cm}{\textbf{Classical communication}} &\multirow{2}{6cm}{All participants reveal their choice of basis for each run.} & \multirow{2}{6cm}{All participants reveal their choice of basis for each run.}\\
&&\\
&\multirow{2}{6cm}{The rounds - in which the computational basis is chosen - are kept.}&\multirow{2}{6cm}{The rounds - in which the same basis choice is made - are kept.}\\
&&\\
\hline

\multirow{2}{3cm}{\textbf{Eavesdropping check}} &\multirow{2}{6cm}{The data of a subset of rounds is revealed to detect eavesdropping.} & \multirow{2}{6cm}{The data of a subset of rounds is revealed to detect eavesdropping.}\\
&&\\
\hline

\multirow{5}{3cm}{\textbf{Key generation}} &\multirow{2}{6cm}{The rest of the rounds constitute key symbols.} & \multirow{2}{6.5cm}{The rounds in which all participants of a given layer choose the same basis- constitute keys. }\\
&&\\
\cline{2-3}

&\multicolumn{2}{c|}{\multirow{3}{14cm}{Each participant expresses his/her outcomes in binary representation, i.e., \( o = 2o_1 + o_0 \equiv (o_1o_0) \). The symbols at the unit’s place (i.e, \(o_0\)) constitute a key in layer \( L_2 \) and those at the two’s place (i.e., \(o_1\)) correspond to key symbols in layer \( L_1 \).} }\\
&\multicolumn{2}{c|}{}\\
&\multicolumn{2}{c|}{}\\
\hline

\end{tabular}
}
\end{center}
\caption{Comparison of LQKD Protocols with reducible multi-dimensional entangled and separable states as resources}
\label{OAM}
\end{table}
The protocols shown in table (\ref{OAM}) have been designed keeping OAM states of light in mind.

For the aims of this paper, it is sufficient for us to focus our attention on the simplest network containing only three participants, {\it viz.}, Alice, Bob$_1$, and Bob$_2$, divided into two layers, $L_1$ and $L_2$. The first layer $L_1$ consists of two participants (Alice and Bob$_1$) and the second layer $L_2$ consists of all three of them. We first present layered quantum key distribution protocols employing multidimensional OAM states both entangled and separable in the table (\ref{OAM}). Of particular importance for us is the fact that the resource state $|\Psi\rangle$ employed in protocol I and the states belonging to bases $S_1$ and $S_2$ are all reducible, i.e., they can be written as tensor products of multiqubit states\footnote{$|\Psi\rangle\equiv\frac{1}{2}\Big\{\Big(|00\rangle+|22\rangle\Big)|0\rangle+\Big(|11\rangle+|33\rangle\Big)|1\rangle\Big\}_{AB_1B_2}=\frac{1}{2}\Big(|00\rangle+|11\rangle\Big)_{AB_1}\Big(|000\rangle+|111\rangle\Big)_{AB_1B_2}.$\\
$S_1: \{|00\rangle, |11\rangle, |20\rangle, |31\rangle\}\equiv \{|000\rangle, |011\rangle, |100\rangle, |111\rangle\}$\\
$S_2: \{|0'+\rangle, |1'-\rangle, |2'+\rangle, |3'-\rangle\equiv |+++\rangle, |+--\rangle, |-++\rangle, |---\rangle\}$}.
\begin{table}[hbt!]
\centering
\begin{tabular}{|c|c|c|}
\hline

\multirow{2}{*}{\textbf{Task}} &\multicolumn{2}{c|}{\multirow{2}{*}{\textbf{Key distribution in layered networks}}}\\
&\multicolumn{2}{c|}{}\\
\hline

\multirow{4}{*}{\textbf{Features $\downarrow$}} &&\\
&\textbf{Protocol III} & \textbf{Protocol IV}\\
& \multirow{1}{*}{\textbf{(with entangled qubits)}} & \multirow{1}{*}{\textbf{(with separable qubits)}}\\
&&\\
\hline

\multirow{4}{*}{\textbf{Resource states}} & \multirow{2}{*}{$\ket{\psi_1}^{(L_1)}\equiv\frac{1}{\sqrt{2}}\Big(\ket{00}+\ket{11}\Big), 
$} & \multirow{2}{*}{${S}_1^{(L_1)}:\Big\{\ket{00},\ket{11}\Big\}, { S}_2^{(L_1)}: \Big\{\ket{++},\ket{--}\Big\}; 
$}\\
&&\\
&\multirow{2}{*}{$\ket{\psi_2}^{(L_2)}\equiv\frac{1}{\sqrt{2}}\Big(\ket{000}+\ket{111}\Big)$}&${ S}_1^{(L_2)}:\Big\{\ket{000},\ket{111}\Big\}, { S}_2^{(L_2)}: \Big\{\ket{+++},\ket{---}\Big\}$\\
&&\\
\hline
\multirow{4}{3cm}{\textbf{Initialisation by Alice}} &\multirow{4}{6cm}{Keeps the first subsystem from both states with herself. Sends the second and third subsystems to Bob\(_1\) and Bob\(_2\) respectively.} & \multirow{4}{6cm}{ Randomly prepares a state from \(S_1\) or \(S_2\) for both layers and sends the first and the second subsystems to Bob\(_1\) and Bob\(_2\) respectively.}\\
&&\\
&&\\
&&\\
\hline

\multirow{5}{3cm}{\textbf{Operations}} &\multirow{3}{6cm}{Every participant measures incoming qubits in the $X$ (diagonal/antidiagonal) or $Z$ ($|H\rangle/|V\rangle$) bases. This process is iterated many times. } & \multirow{3}{6cm}{Every participant measures incoming qubits in the $X$ (diagonal/antidiagonal) or $Z$ bases ($|H\rangle/|V\rangle$). This process is iterated many times. }\\
&&\\
&&\\
&&\\
\hline

\multirow{4}{3cm}{\textbf{Classical communication}} &\multirow{2}{6cm}{All participants reveal their choice of basis for each run.} & \multirow{2}{6cm}{All participants reveal their choice of basis for each run.}\\
&&\\
&\multirow{2}{6cm}{The rounds - in which the \textbf{computational basis} is chosen - are kept.}&\multirow{2}{6cm}{The rounds - in which \textbf{the same basis} choice is made - are kept.}\\
&&\\
\hline

\multirow{2}{3cm}{\textbf{Eavesdropping check}} &\multirow{2}{6cm}{The data of a subset of rounds is revealed to detect eavesdropping.} & \multirow{2}{6cm}{The data of a subset of rounds is revealed to detect eavesdropping.}\\
&&\\
\hline

\multirow{2}{3cm}{\textbf{Key generation}} &\multirow{2}{6cm}{The rest of the rounds constitute key symbols. } & \multirow{2}{6.5cm}{The rounds in which all participants of a given layer choose the same basis- constitute keys. }\\
&&\\
\hline

\end{tabular}
\caption{Comparison of LQKD Protocols with multiqubit entangled and separable states}
\label{qubit}
\end{table}
Thereafter, we present an equivalent  LQKD protocol that accomplishes the same task of QKD in layered networks by employing entangled two-qubit Bell states and  three-qubit GHZ states (table (\ref{qubit})). The last protocol shows that the task of QKD protocols in layered networks can be implemented with separable qubits as well (table (\ref{qubit})).

Since layered QKD protocols can be implemented with polarisation qubits as well,  multidimensional entangled OAM states do not offer any intrinsic advantage in the task of QKD in layered quantum networks.  The generation yield of multidimensional entangled OAM states, which is used as a resource state in the first protocol ($\sim 10^{-3}-10^{-1}$ Hz) (see, for example, \cite{erhard2020advances} and references therein). The yield of separable OAM states is of the order of $10-10^3$ Hz. This is lower  than the yield of heralded single-photon  generation (in the polarization domain) from crystals ($\sim 10^3-10^5$ per second) which is used as a resource in the last protocol \cite{Mosley08, Ljunggren05, Scholz09, fedrizzi2007wavelength}. Due to the lower yield of multidimensional OAM entangled states, implementations of QKD protocols with qubit states are much more feasible in the current state of technology as compared to those with multidimensional OAM entangled states. The generation yields  of two-party photonic entangled states in different degrees of freedom (e.g., spatial, time-bin, frequency) are also much higher  ($\sim 10^3-10^6$ pairs per second) \cite{kim2017two} than that of multidimensional entangled OAM states  ($\sim 10^{-3}-10^{-1}$ pairs per second).

This leaves us with a question of interest in this paper: {\it are there secure quantum communication tasks that inevitably require multidimensional states as a resource? What we mean to say is that these tasks cannot be implemented with several copies of entangled multiparty states belonging to a Hilbert space of lower dimensions.} To answer this question, we start by showing that OAM states are quintessential in increasing the key generation rate. Thereafter, we present different tasks  for which OAM states are absolutely essential.
\section*{Enhancement in key generation rates: indispensability of OAM states}
Consider the same network \( L \) with three participants: Alice, Bob$_1$, and Bob$_2$, spread across two layers, $L_1$ and $L_2$. $L_1$ has only Alice and Bob$_1$, while $L_2$ encompasses all three participants. Suppose that for layer \( L_1 \), our aim is to achieve a sifted key rate of \( \log_2d \) bits per transmission, where \( d \neq 2^n\). In order to realize this, qudits of dimension \( d \) need to be transmitted. This requires the employment of at least two bases: the computational (OAM) basis represented by \( S_1 \) as
\begin{align}
 S_1 \equiv \{|0\rangle, |1\rangle, \cdots, |d-1\rangle\} 
 \end{align}
and the Fourier (angular) basis \( S_2 \) given by 
\begin{align}
S_2\equiv\Big\{\frac{1}{\sqrt{d}}\sum_{j=0}^{d-1}\omega^{ij}|j\rangle, 0\leq i\leq d-1\Big\}.
\end{align}
Given that OAM states are physical candidates for the implementation of  qudits, they become indispensable when the objective is to enhance the key generation rate in a layer relative to qubits.

\section*{Tasks implementable with irreducible multidimensional entangled OAM states only}
\hypertarget{Tasks implementable with irreducible multidimensional entangled OAM states only}{}
Our focus is on the tasks that may be implemented with irreducible multidimensional entangled OAM states. These protocols bring out the true potential of OAM states in quantum communication. It is because even if these protocols were to be implemented with multiqubit polarisation states,  we shall see that each party has to hold a large number  of qubits and has to perform entangled states measurements-- none of which is required for the multidimensional irreducible entangled OAM states. 
\subsection*{Controlled - secret sharing and key distribution (c-SSKD)}
\hypertarget{controlled-secret sharing and key distribution}{}
In the following, as an application, we show that using high-dimensional states, keys, and secrets can be shared simultaneously in a controlled manner. For this purpose, consider a network of three participants namely Alice, Bob, and Charlie. Suppose that Charlie is the participant who is given the control to decide in which rounds keys/secrets are to be distributed.
For this purpose, we employ the \textit{irreducible multidimensional} OAM states (up to a normalization constant) as the resource state, 
\begin{align}    |\psi^{(4)}\rangle\equiv\sum_{j=0}^3|u_j\rangle|j\rangle|u_j\rangle-2|333\rangle,
\end{align}
where $\{\ket{u_j}\}$ are mutually unbiased with states in the computational basis $\{\ket{j}\}$ and are defined as follows:
\begin{align}
    |u_0\rangle &\equiv \frac{1}{2}\Big(|0\rangle+|1\rangle+|2\rangle+|3\rangle\Big),~~ |u_1\rangle \equiv \frac{1}{2}\Big(|0\rangle-|1\rangle+|2\rangle-|3\rangle\Big)\nonumber\\
    |u_2\rangle &\equiv \frac{1}{2}\Big(|0\rangle+|1\rangle-|2\rangle-|3\rangle\Big),~~
    |u_3\rangle \equiv \frac{1}{2}\Big(|0\rangle-|1\rangle-|2\rangle+|3\rangle\Big).
\end{align}
The steps of the protocol to realize the task are as follows:

 \begin{wrapfigure}{r}{6.5cm}
\centering
    \includegraphics[scale=0.25]{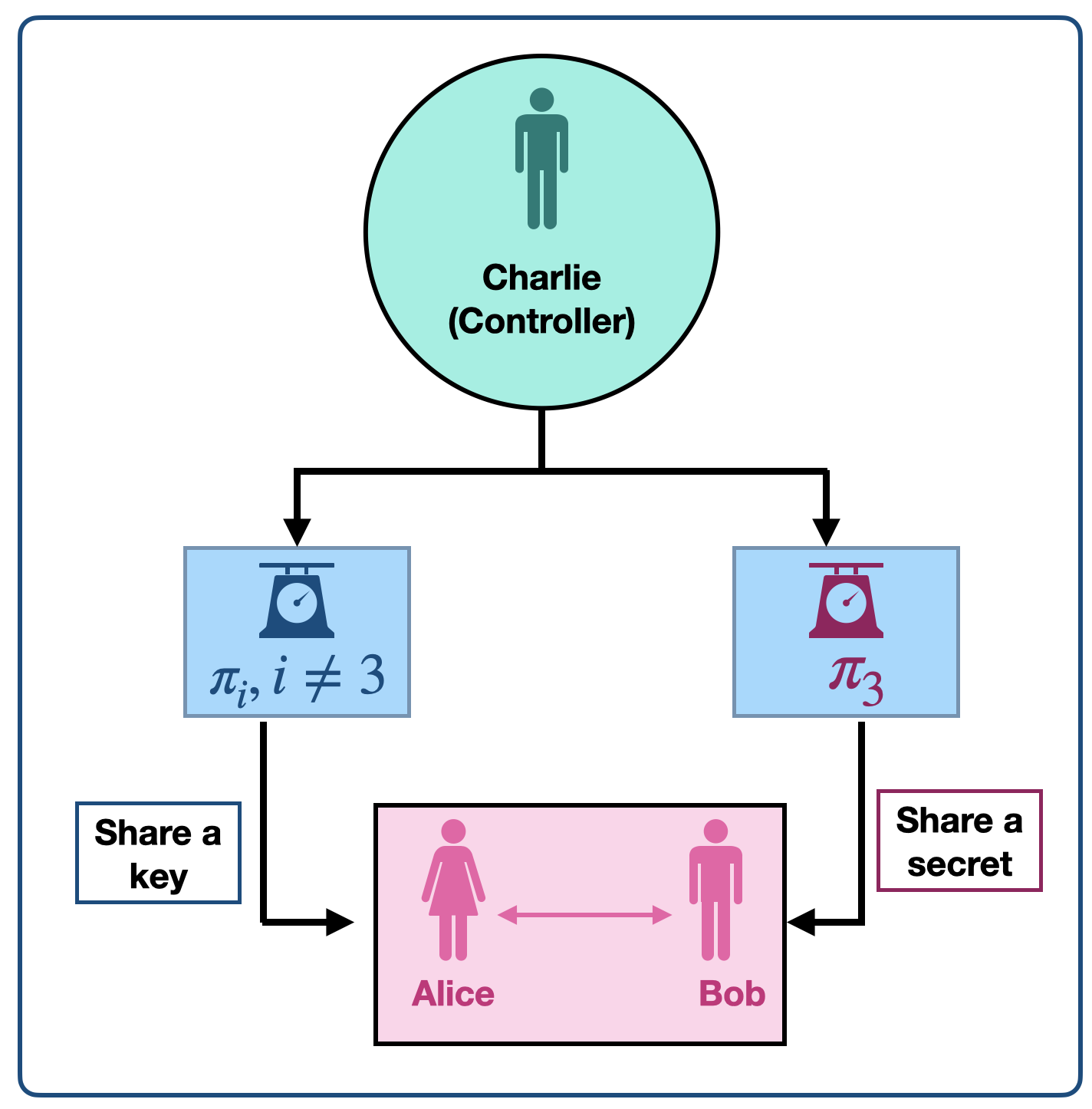}
    \caption{Pictorial description of the task implemented with irreducible multidimensional entangled states with specific basis choice.}
    \label{fig:ckd}
\end{wrapfigure} 
\begin{enumerate}
    \item \textbf{State distribution:} Suppose that Charlie prepares the state $\ket{\Psi^{(4)}}$. He sends the first and the third subsystems to Alice and Bob keeping the second with himself.
    \item \textbf{Control operations:} Charlie performs one of the three operations depending upon the task he wishes as follows:
    \begin{enumerate}
        \item Charlie performs a projective measurement $\pi_3=\ket{3}\bra{3}$ on his subsystem. He wishes to utilize the round for secret sharing.  For the purpose of eavesdropping check, she also sends some Decoy pulses.
        \item He performs one of the projective measurements $\{\pi_i=\ket{i}\bra{i},~i\in 0,1,2\}$ whenever he wishes to share a key with Alice and Bob.
        \item Charlie also measure in the conjugate basis $\{\ket{u_j}\}$ in order to ensure robustness against eavesdropping.
    \end{enumerate}
    \item Alice and Bob randomly and independently choose to measure either in the computational basis or in the conjugate basis.
    \item Steps $(1-3)$ constitute one round and are repeated for a large number of rounds.
    \item Afterward, Alice and Bob reveal their choice of basis for each round. Charlie reveals the rounds in which he measures in the conjugate basis or  $\pi_3$ or from the rest of the measurement.
    \item The data of a subset of the rounds in which Decoy states have correlations are employed to check for the presence of eavesdropping.
    \item  In the absence of an eavesdropper, the data of the rest of the rounds in which Charlie measures $\pi_3$, and Alice and Bob measure in the same basis i.e., either computational or conjugate basis shares a secret between Alice and Bob following the rule $s=a\oplus_4 b$, where $a,b$ represent the outcome of Alice and Bob and symbol $\oplus_4$ represent addition modulo $4$.
    \item The data of the rest of the rounds in which Charlie measures the rest of the projections, and Alice and Bob measure in the conjugate bases- shares a key among all three participants. In fact, the measurement of Charlie decides the key symbol which is shared in the corresponding round.
\end{enumerate}
In this manner, Charlie shares either a key of his choice or allows Alice and Bob to share a secret. The same has been shown pictorially in Figure (\ref{fig:ckd}). This particular feature is possible, thanks to the existence of irreducible high-dimensional states.

Please note that in the key generation rounds, three key symbols are generated with an equal probability whereas secret sharing results in the generation of $4$ secret symbols with equal probability.

\subsection*{Basis-dependent simultaneous  secret sharing and key distribution (bd-SSSKD)}
\hypertarget{Basis-dependent simultaneous secret sharing and key distribution}{}
In the previous section, we have shown how irreducible high-dimensional states allow for the controlled distribution of keys as well as secrets using the single resource states on the free will of the controller `Charlie'. However, this is not the only task that is implementable with such states. In fact, if one harnesses the multi-dimensional states as well, one can implement various other tasks simultaneously in a given network. 

\begin{wrapfigure}{r}{7.1cm}
\centering
    \includegraphics[scale=0.22]{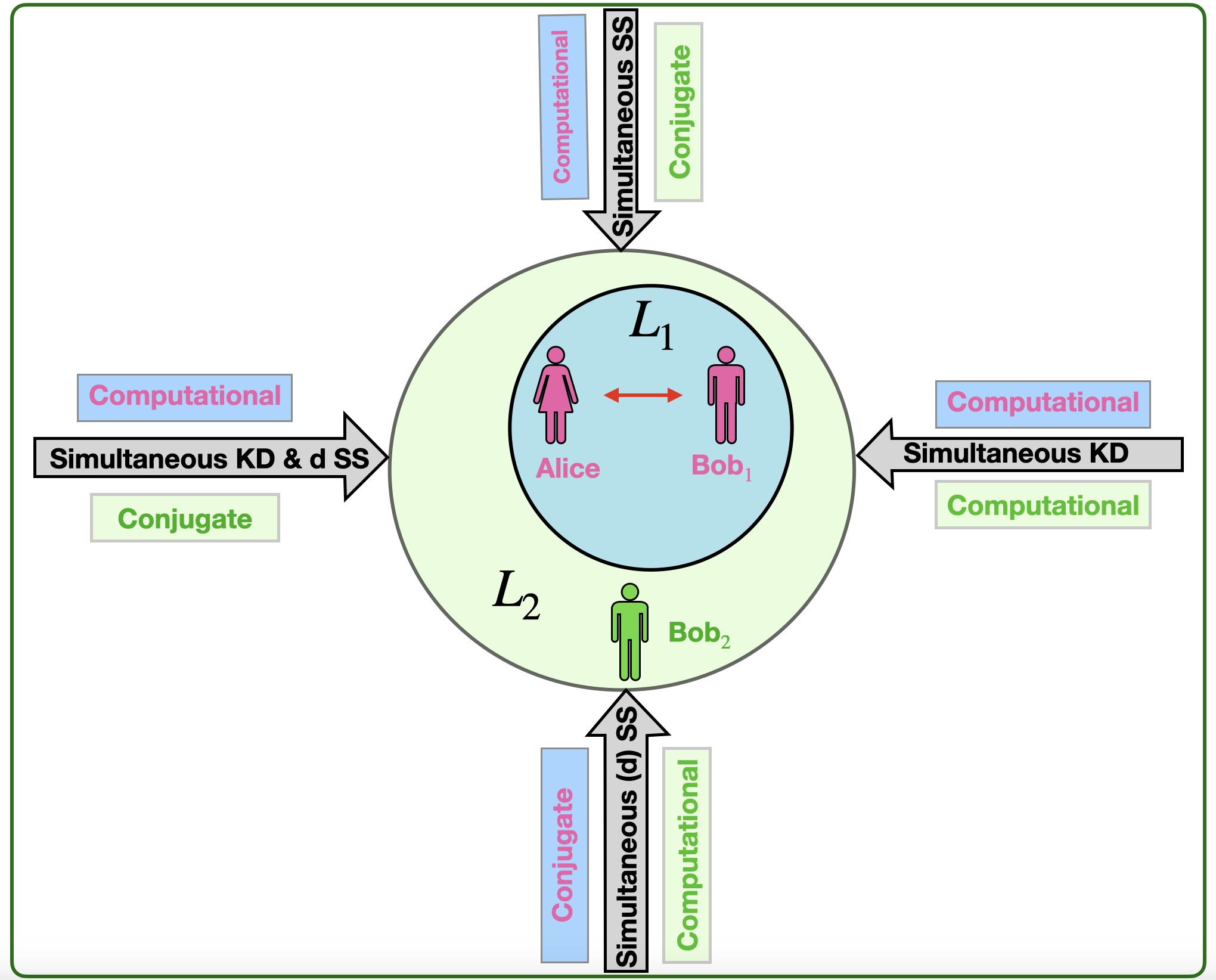}
    \caption{Pictorial description of the task implemented with irreducible multidimensional entangled states with specific basis choice.}
    \label{fig:sbd}
\end{wrapfigure}

To show that this indeed is the case, we consider a network of three participants, say, Alice, Bob$_1$, and Bob$_2$. Let these participants be distributed in the two layers $L_1$ and $L_2$. Layer $L_2$ consists of all three participants while layer $L_1$ consists of only Alice and Bob$_1$ as also shown in Figure \ref{fig:sbd}.

In the following, we show that by employing irreducible multi-dimensional states, one can implement (i) simultaneous key distribution in the two layers, (ii) Simultaneous sharing of a key and a deterministic secret in the network, and (iii) sharing of secrets in both layers and, (iv) simultaneous (deterministic) secret sharing-- all four tasks with a single state. Please note that these tasks correspond to a different choice of bases which is required to ensure the robustness of the protocol against eavesdropping. Thus, the following protocol utilizes the resource states in its full form. For this purpose, we utilize the following resource state:
\begin{align}
    \ket{\Psi}=\frac{1}{2}\Big(\ket{000}+\ket{111}+\ket{220}-\ket{331}\Big).
\end{align}

 \begin{center}
     \textbf{The protocol}
 \end{center}
 The steps of the protocol are as follows:
 \begin{enumerate}
\item \textit{Resource distribution:} Suppose that Alice, Bob$_1$, and Bob$_2$ are provided with the first, the second, and the third subsystem of the state $\ket{\psi}$.
\item \textit{Measurements:} All three participants randomly and independently choose to measure either in the computational (OAM) or in the conjugate (angular) basis on their respective subsystems.
\item Steps $(1-2)$ constitute one round and are repeated for a large number of rounds.
\item Afterward, Alice, Bob$_1$, and Bob$_2$ reveal their choice of basis for each round on an authenticated classical channel\footnote{An authenticated classical channel is the one on which Eve can listen to all the shared information, however cannot tamper the information. }.
\item \textit{Eavesdropping check:} The data of the subset of the rounds are used to check for the presence of an eavesdropper.
\item In the absence of an eavesdropper, the data of the rounds in which Alice and Bob$_1$ choose the same basis are used to generate keys and secrets as discussed below.
   \end{enumerate}
Since Alice and Bob$_1$  are in the hold of four-level subsystems, they generate $4$ symbols. To achieve simultaneous distribution of keys/secrets, they write their outcomes in the binary representation. Suppose that outcomes of Alice, Bob$_1$, and Bob$_2$ are represented by $a,~b_1,$ and $b_2$ respectively. In the binary representation, they take the form:
\begin{align}
    a\equiv a^{(1)}a^{(0)},~~b_1\equiv b_1^{(1)}b_1^{(0)}
\end{align}
\noindent\textbf{Simultaneous key distribution (KD):} The rounds in which all three participants measure in the computational basis are used to share keys in the two layers. This can be seen directly from table \ref{tab:skd}. Key symbols in blue color share key in the layer $L_1$ while the symbols in red color share key in the layer $L_2$. The confidentiality of the keys being shared in the two layers is also ensured. The same is also reflected in the table as Bob$_2$ who is not the participant of layer $L_1$ cannot infer what is being shared (in blue color).

\begin{table}[h!]
    \centering
    \begin{tabular}{|c|c|c|c|c|}
    \hline
    \rowcolor{lightgray}    \multicolumn{2}{|c}{Alice} &\multicolumn{2}{|c|}{Bob$_1$} &   Bob$_2$ \\\hline
    \rowcolor{lightgray}   Outcome&Binary representation& Outcome& Binary representation   & Outcome \\\hline
     $0$ & {\color{blue}$0$}{\color{red}$0$} & $0$ & {\color{blue}$0$}{\color{red}$0$}  & {\color{red}$0$}\\
    \rowcolor{lightgray}   $1$ &{\color{blue} $0$}{\color{red}$1$} & $1$ & {\color{blue}$0$}{\color{red}$1$} & {\color{red}$1$}\\
       $2$ &{\color{blue} $1$}{\color{red}$0$} & $2$ & {\color{blue}$1$}{\color{red}$0$} & {\color{red}$0$}\\
    \rowcolor{lightgray}   $3$ & {\color{blue}$1$}{\color{red}$1$} & $3$ & {\color{blue}$1$}{\color{red}$1$} & {\color{red}$1$}\\\hline
    \end{tabular}
    \caption{Correlations in the outcome of Alice, Bob$_1$, and Bob$_2$ in the computational basis}
    \label{tab:skd}
\end{table}

\noindent\textbf{Simultaneous secret sharing (SS):} When all three participants measure in the conjugate basis, secrets are being shared in the two layers. This is because for the measurement in the conjugate basis, all three participants will reveal all possible combinations of their outcomes resulting in a total of $32$ possibilities. As before, Alice and Bob$_1$ write their outcomes in binary representation. Following which secrets in two layers are,
\begin{align}
    s_1=a^{(1)}\oplus b_1^{(1)},~~~~~~s_2=a^{(0)}\oplus b_1^{(0)}\oplus b_2.
\end{align}

\noindent\textbf{Simultaneous key distribution (KD) and deterministic secret sharing (d SS) :} The rounds in which Alice and Bob$_1$ choose the computational basis and Bob$_2$ chooses the conjugate basis are employed to share both keys and secrets simultaneously. The same can be seen from table \ref{tab:skdss1}. Please note that symbols at the unit's $(2^0 s)$ (in blue color) place share a key in layer $L_1$ and those of at $2^1 s$ place (in red color) share a secret in layer $L_2$ as $s=a^{(0)}\oplus b_1^{(0)}=b_2$. The symbol $\oplus$ represents addition modulo $2$.

\begin{table}[h!]
    \centering
    \begin{tabular}{|c|c|c|c|c|}
    \hline\rowcolor{lightgray}
       \multicolumn{2}{|c|}{\textcolor{black}{Alice}} &\multicolumn{2}{|c|}{\textcolor{black}{Bob$_1$}} & \textcolor{black}{Bob$_2$} \\
    \hline
    \rowcolor{lightgray}
      Outcome & Binary representation & Outcome & Binary representation & Outcome \\
    \hline\hline
    \rowcolor{white}
     $0$ & {\color{blue}$0$}{\color{red}$0$} & $0$ & {\color{blue}$0$}{\color{red}$0$} & {\color{red}$0$} \\
    \rowcolor{lightgray}
     $1$ & {\color{blue}$0$}{\color{red}$1$} & $1$ & {\color{blue}$0$}{\color{red}$1$} & {\color{red}$0$} \\
    \rowcolor{white}
     $2$ & {\color{blue} $1$}{\color{red}$0$} & $2$ & {\color{blue}$1$}{\color{red}$0$} & {\color{red}$0$} \\
    \rowcolor{lightgray}
     $3$ & {\color{blue} $1$}{\color{red}$1$} & $3$ & {\color{blue}$1$}{\color{red}$1$} & {\color{red}$0$} \\
     \hline
    \rowcolor{white}
     $0$ & {\color{blue} $0$}{\color{red}$0$} & $1$ & {\color{blue}$0$}{\color{red}$1$} & {\color{red}$1$} \\
    \rowcolor{lightgray}
     $1$ & {\color{blue}$0$}{\color{red}$1$} & $0$ & {\color{blue}$0$}{\color{red}$0$} & {\color{red}$1$} \\
    \rowcolor{white}
     $2$ & {\color{blue}$1$}{\color{red}$0$} & $3$ & {\color{blue} $1$}{\color{red}$1$} & {\color{red}$1$} \\
    \rowcolor{lightgray}
     $3$ & {\color{blue}$1$}{\color{red}$1$} & $2$ & {\color{blue}$1$}{\color{red}$0$} & {\color{red}$1$} \\
    \hline
    \end{tabular}
    \caption{Correlations in the outcome of Alice, and Bob$_1$ choose the computational basis, and Bob$_2$ choose the conjugate basis}
    \label{tab:skdss1}
\end{table}

\noindent\textbf{Simultaneous (deterministic) secret sharing ((d) SS) :} The rounds in which Alice and Bob$_1$ choose the conjugate basis and Bob$_2$ chooses the computational basis are employed to share secrets in two layers. However, a secret shared in the layer $L_2$ is known to Bob$_2$. That is to say, collaboration is required only between Alice and Bob$_1$ and so the name deterministic secret sharing. The same can be seen from table \ref{tab:skdss2} in which symbols in red constitute secret in layer $L_1$ as $s_1\equiv a^{(0)}\oplus b_1^{(0)}$. However, the symbols in blue are used to generate a secret in the layer $L_2$ as $s_2\equiv a^{(1)}\oplus b_1^{(1)}=b_2$.

\begin{table}[h!]
    \centering
    \begin{tabular}{|c|c|c|c|c|}
    \hline
    \rowcolor{lightgray}
       \multicolumn{2}{|c|}{Alice} &\multicolumn{2}{|c|}{Bob$_1$} & Bob$_2$ \\
    \hline
    \rowcolor{lightgray}
      Outcome & Binary representation & Outcome & Binary representation & Outcome \\
    \hline\hline
     $0$ & {\color{blue}$0$}{\color{red}$0$} & $0$ & {\color{blue}$0$}{\color{red}$0$} & {\color{blue}$0$} \\
     \rowcolor{lightgray} $0$ & {\color{blue}$0$}{\color{red}$0$} & $1$ & {\color{blue}$0$}{\color{red}$1$} & {\color{blue}$0$} \\
     $1$ & {\color{blue} $0$}{\color{red}$1$} & $0$ & {\color{blue}$0$}{\color{red}$0$} & {\color{blue}$0$} \\
    \rowcolor{lightgray}  $1$ & {\color{blue} $0$}{\color{red}$1$} & $1$ & {\color{blue}$0$}{\color{red}$1$} & {\color{blue}$0$} \\
     $2$ & {\color{blue}$1$}{\color{red}$0$} & $2$ & {\color{blue}$1$}{\color{red}$0$} & {\color{blue}$0$} \\
    \rowcolor{lightgray}  $2$ & {\color{blue}$1$}{\color{red}$0$} & $3$ & {\color{blue}$1$}{\color{red}$1$} & {\color{blue}$0$} \\
     $3$ & {\color{blue} $1$}{\color{red}$1$} & $2$ & {\color{blue}$1$}{\color{red}$0$} & {\color{blue}$0$} \\
    \rowcolor{lightgray}  $3$ & {\color{blue} $1$}{\color{red}$1$} & $3$ & {\color{blue}$1$}{\color{red}$1$} & {\color{blue}$0$} \\
     \hline
     $0$ & {\color{blue} $0$}{\color{red}$0$} & $2$ & {\color{blue}$1$}{\color{red}$0$} & {\color{blue}$1$} \\
    \rowcolor{lightgray}  $0$ & {\color{blue}$0$}{\color{red}$0$} & $3$ & {\color{blue}$1$}{\color{red}$1$} & {\color{blue}$1$} \\
     $1$ & {\color{blue}$0$}{\color{red}$1$} & $2$ & {\color{blue} $1$}{\color{red}$0$} & {\color{blue}$1$} \\
    \rowcolor{lightgray}  $1$ & {\color{blue}$0$}{\color{red}$1$} & $3$ & {\color{blue}$1$}{\color{red}$1$} & {\color{blue}$1$} \\
     $2$ & {\color{blue} $1$}{\color{red}$0$} & $1$ & {\color{blue}$0$}{\color{red}$1$} & {\color{blue}$1$} \\
    \rowcolor{lightgray}  $3$ & {\color{blue}$1$}{\color{red}$1$} & $0$ & {\color{blue}$0$}{\color{red}$0$} & {\color{blue}$1$} \\
     $3$ & {\color{blue}$1$}{\color{red}$1$} & $1$ & {\color{blue} $0$}{\color{red}$1$} & {\color{blue}$1$} \\
   \rowcolor{lightgray}   $2$ & {\color{blue}$1$}{\color{red}$0$} & $0$ & {\color{blue}$0$}{\color{red}$0$} & {\color{blue}$1$} \\
    \hline
    \end{tabular}
    \caption{Correlations in the outcomes when Alice and Bob$_1$ choose the conjugate basis, and Bob$_2$ choose the computational basis. In this particular round, Bob$_2$ based on his outcome knows the secret that is being shared between Alice and Bob$_1$. However, the symbols at the unit's place are completely independent of the measurement outcome of Bob$_2$. So, this particular round allows the sharing of two secrets between Alice and Bob$_1$- one which is completely deterministic to Bob$_2$ and the other which is completely unknown to Bob$_2$. }
    \label{tab:skdss2}
\end{table}

In this manner, using an irreducible multi-dimensional state, all four kinds of tasks are realized. The tasks and the corresponding choices of basis are pictorially shown in Figure (\ref{fig:sbd}). In all the layers and tasks, binary key symbols are generated with equal probability leading to $1$ bit of sifted key/secret generation rate.

These protocols are characterized by the local dimensionality of each subsystem and the number of parties. For distributing a secret and a key among different participants of layered networks,  multipartite multidimensional irreducible entangled states are required. The number of parties should be equal to the number of participants in a given network. Depending on the nature of irreducible entanglement, different kinds of quantum key distribution and quantum secret-sharing protocols may be realized. Notably, the number of these tasks is much more as compared to layered quantum key distribution. It is because the latter task involves bispearable states (belonging to the Hilbert space of a given dimension). On the other hand, the former tasks involve irreducible entangled states.

Both the protocols c-SSKDP and bd-SSSKDP employ irreducible entangled states as resources to implement the tasks. The robustness of these protocols, as in the conventional entanglement-based protocols, is ensured by the property that entangled states exhibit correlations in more than one basis. Any interventions by Eve would inevitably disturb these correlations exhibited by the subsystems, resulting in the detection of eavesdropping. A rigorous information-theoretic security analysis constitutes an interesting study that will be taken up elsewhere.

\section*{Conclusion}
\hypertarget{Conclusions}{}
In this paper, we have shown that the task of layered quantum key distribution may be implemented with both reducible entangled OAM states and several copies of entangled/separable multiqubit states distributed in different layers. For this reason, reducible entangled OAM states do not offer any intrinsic advantage in this task. However, there are integrated quantum communication tasks-- for which OAM states are absolutely essential. We have identified two such secure quantum communication tasks-- (i) controlled secret sharing and key distribution and, (ii) basis-dependent simultaneous secret sharing and key distribution-- that can be implemented with high-dimensional irreducible entangled OAM states only. It shows that multidimensional entangled OAM states offer new possibilities in secure quantum communication. We believe that this study opens up a new direction in which multidimensional irreducible entangled OAM states act as resources. Analogous in-depth investigations can be undertaken in interconnected domains such as quantum computation, quantum-based search algorithms, and quantum sensing.

\section*{Acknowledgements}
We would like to express our sincere gratitude to V. Ravishankar for motivating us to start the project and for his valuable contribution to the ideas that greatly influenced the direction of this research. His insights were integral to the development of this paper. 

\section*{Author contribution statement}
Both the authors contributed equally in all respects.

\section*{Data availability statement}
No data was generated during this study.

\noindent\section*{Competing interests} 
The authors declare no competing interests.

\end{document}